\documentclass[aps,prl,reprint,groupedaddress]{revtex4-2}

\usepackage[utf8]{inputenc}
\usepackage{latexsym, graphicx} 
\usepackage{amsmath,amsfonts,amssymb}
\usepackage{cancel}
\usepackage{mathrsfs,dsfont}
\usepackage{tensor}
\usepackage{xcolor}
\usepackage[shortlabels]{enumitem}
\usepackage{hyperref}
\hypersetup{
	colorlinks,%
	citecolor=blue,%
	filecolor=blue,%
	linkcolor=blue,%
	urlcolor=blue,
	linktoc=page
}

\newcommand*{\scri}{\ensuremath{\mathscr{I}}}

\newcommand*{\dd}{\mathop{}\!d}
\newcommand*{\x}{\mathbf{x}}

\begin{document}
	\title{Soft Theorems and Spontaneous Symmetry Breaking}
	\author{Shreyansh Agrawal}
	\email{sagrawal@sissa.it}
	
	\author{Kevin Nguyen}
	\email{kevin.nguyen2@ulb.be}
	
	\affiliation{\vspace{0.2cm} ${}^\dagger$Universit\'e Libre de Bruxelles and International Solvay Institutes, ULB-Campus Plaine
		CP231, 1050 Brussels, Belgium}
	
	\affiliation{\vspace{0.2cm} ${}^*$SISSA, Via Bonomea 265, 34136 Trieste, Italy}
	\affiliation{INFN, Sezione di Trieste, Via Valerio 2, 34127, Italy}

	\begin{abstract}
		The soft photon and soft graviton theorems of Weinberg are known to derive from conservation laws associated with asymptotic symmetries. Within the corresponding classical theories, one often speaks of spontaneous symmetry breaking and vacuum degeneracy, but a genuine quantum description of this phenomenon has largely been lacking. Here we establish spontaneous breaking of asymptotic symmetries and the existence of Goldstone `particles' using exclusively the language of quantum field theory. This is made possible through the reformulation of massless scattering theory in terms of carrollian conformal field theory, and the observation that soft theorems correspond to Ward identities of broken symmetries. A suitable version of Goldstone theorem shows that there must exist zero-momentum particles described by conformal fields on the celestial sphere, in agreement with the common lore. More specifically, these belong to unitary representations in the discrete series of the Lorentz group, and are therefore naturally equipped with logarithmic two-point functions. We discuss the relevance of these observations to the problem of infrared divergences that scattering amplitudes suffer from. 
	\end{abstract}
	\maketitle
	
	\section{Introduction}
	\label{Intro}
	The soft photon and soft graviton theorems satisfied by scattering amplitudes constitute an important milestone in the study of quantum electrodynamics and gravity \cite{Weinberg:1965nx}. Indeed, the infrared divergences which they imply at loop-level are symptomatic of the inadequacy of conventional scattering theory with massless gauge particles, even though the use of inclusive cross sections has allowed physicists to retain their predictivity \cite{Bloch:1937pw,Yennie:1961ad,Kinoshita:1962ur,Lee:1964is}. The causes of this dramatic failure are the long-range interactions carried by these massless particles, which invalidate the underlying assumption that asymptotic particle states propagate freely. The resulting asymptotic acceleration of the charged particles does not vanish fast enough, resulting in infrared divergences \cite{Zwanziger:1973mp,Sahoo:2018lxl,Compere:2025tzr}.
	
	More recently a new perspective on this subject has emerged. Indeed it has been understood that soft theorems are manifestations of \textit{asymptotic symmetries} \cite{Strominger:2013lka,Strominger:2013jfa,He:2014laa,He:2014cra,Kapec:2015ena,Campiglia:2015qka,Strominger:2017zoo}, and that they are the scattering counterpart of the memory effects studied within the classical theories \cite{Zeldovich:1974gvh,Braginsky:1987aa,Christodoulou:1991cr,Bieri:2013hqa,Strominger:2014pwa}. Within the classical theories asymptotic symmetries are further viewed as transitions between degenerate `vacua', suggesting that they are spontaneously broken. The degenerate vacuum solutions are distinguished by some free functions on the celestial sphere that behave as conformal fields under the Lorentz group, and which are naturally called \textit{Goldstone modes} as they shift under the corresponding asymptotic symmetries. Eventually promoting these Goldstone modes to quantum conformal fields, it has been argued that they effectively account for the infrared divergences in scattering amplitudes \cite{Nande:2017dba,Himwich:2020rro,Nguyen:2021ydb,Donnay:2022hkf,Agrawal:2023zea,Kapec:2021eug,He:2024skc}.
	
	Nonetheless it is fair to say that the spontaneous symmetry breaking of asymptotic symmetries, as understood thus far, has not been given a proper quantum field theory treatment. The reason for this is simply that asymptotic symmetries look very exotic from the perspective of local quantum field theory in Minkowski spacetime. In fact, we generally do not even know how to describe their action away from the asymptotic boundaries of spacetime. In this letter we will remedy this situation by giving up on a Minkowskian field theory description altogether, and rather focus on the `holographic' description in terms of local \textit{carrollian} fields defined at null infinity \cite{Bagchi:2016bcd,Banerjee:2018gce,Banerjee:2019prz,Donnay:2022aba,Bagchi:2022emh,Donnay:2022wvx,Nguyen:2023vfz,Nguyen:2023miw,Mason:2023mti,Bagchi:2024gnn,Ruzziconi:2024kzo,Nguyen:2025sqk}. The basic observation behind carrollian holography is simple: massless particles states are carried by carrollian conformal fields living at null infinity, as much as they are carried by relativistic tensor fields in Minkowski spacetime. Henceforth massless scattering amplitudes are equivalent to correlation functions of these carrollian fields. Because asymptotic symmetries act in a transparent way on local carrollian fields, we will be able to describe their spontaneous breaking in a conventional way. 
	
	We organise this letter as follows. After recalling the relation between massless particles and carrollian conformal fields, we describe the action of asymptotic symmetries that underlie Weinberg's soft photon and soft graviton theorems. We then show that the soft theorems can only possibly hold in carrollian field theories wherein the vacuum spontaneously breaks these  asymptotic symmetries. Having established spontaneous breaking of asymptotic symmetries, we prove that it implies the existence of Goldstone modes belonging to \textit{zero-momentum} representations of the Poincar\'e group $\operatorname{ISO}(1,3)$. Zero-momentum fields are conventional conformal fields on the celestial sphere, in agreement with expectations from the (semi)-classical theory. We end with a discussion of the prospects offered by this improved description of the spontaneous breaking of asymptotic symmetries.
	
	\section{Carrollian Ward identities}
	\label{section 2}
	In this section we recall the description of massless scattering theory in terms of a carrollian conformal field theory, broadly referred to as \textit{carrollian holography}. Within this description, we explain how electromagnetic and gravitational asymptotic symmetries are realised, and the Ward identities they imply for carrollian amplitudes.
	
	The carrollian conformal field theory that we will consider in this work mostly concerns massless particles as defined by Wigner, i.e., unitary irreducible representations of the Poincar\'e group $\operatorname{ISO}(1,3)$ with vanishing quadratic Casimir invariant \cite{Wigner:1939cj}. Because the Poincar\'e group is realised as a group of conformal isometries of the carrollian manifold $\scri \approx \mathbb{R} \times S^2$, viewed as the homogeneous space
	\cite{Herfray:2020rvq,Figueroa-OFarrill:2021sxz}
	\begin{equation}
	\scri\simeq \frac{\operatorname{ISO}(1,3)}{(\operatorname{ISO}(2) \ltimes \mathbb{R}^3) \rtimes \mathbb{R}}\,,
	\end{equation}
	one can construct conformal field representations of $\operatorname{ISO}(1,3)$ living on $\scri$ which encode the massless particle states \cite{Banerjee:2018gce,Nguyen:2023vfz}. This manifold is endowed with a conformal equivalence class of degenerate metrics, with standard representative
	\begin{equation}
	\label{scri metric}
	ds^2_{\scri}=0\, \dd u^2+\delta_{ij} \dd x^i \dd x^j\,.
	\end{equation}
	Here $\vec x=x^i$ are cartesian stereographic coordinates on the celestial sphere $S^2$, and we will denote the full set of coordinates by $\x=(u,\vec x)$. Of particular interest is the realisation of~$\scri$ as the future or past component of the null conformal boundary of Minkowski spacetime. See \cite{Nguyen:2022zgs,Nguyen:2020hot} and references therein for a geometrical description of $\scri$ in the context of asymptotically flat gravity. 
	
	The carrollian conformal fields $O_{\Delta,J}(\x)$ which encode massless one-particle states are labeled by the helicity $J$ of the corresponding particle and by some scaling dimension $\Delta$. Their construction as field representations of $\operatorname{ISO}(1,3)$ can be found in \cite{Nguyen:2023vfz,Nguyen:2025sqk}. Here we wish to recall their relation to one-particle operators $a^\dagger_J(p)$. Adopting the momentum parametrisation
	\begin{equation}
	\label{momentum parametrisation}
	p^\mu(\omega, \vec x)=\frac{\omega}{\sqrt{2}}(1+|\vec x|^2,2\vec x,1-|\vec x|^2)\equiv \omega\, q^\mu(\vec x)\,,
	\end{equation}
	the intertwining relation is given through the modified Fourier transform \cite{Banerjee:2018gce,Banerjee:2019prz,Donnay:2022wvx,Nguyen:2023vfz}
	\begin{equation}
	\label{Mellin}
	O_{\Delta,J}(\x)=\int_0^\infty d\omega\, \omega^{\Delta-1} e^{i\omega u}\, a^\dagger_J(p(\omega,\vec x))\,.
	\end{equation}
	This transform can be applied to momentum $\mathcal{S}$-matrix elements $S_n$, thereby defining the \textit{carrollian amplitudes}
	\begin{widetext}
		\begin{equation}
		\label{Mellin amplitudes}
		\langle O^{\eta_1}_{\Delta_1,J_1}(\x_1)\, ...\, O^{\eta_n}_{\Delta_n,J_n}(\x_n) \rangle\equiv \prod_{k=1}^n \int_0^\infty d\omega_k\, \omega^{\Delta_k-1} e^{i \eta_k \omega_k u_k} S_n(1^{J_1}...\,n^{J_n})\,,
		\end{equation}
	\end{widetext}
	where $\eta_k=\pm 1$ depending whether the particle is ingoing (+) or outgoing (-). This is a convention where all particles can be effectively treated as if they were ingoing, with ingoing momenta $p_k^\mu$ as given by \eqref{momentum parametrisation} times $\eta_k$, and ingoing helicity $J_k$. Simply based on the transformation properties of the $\mathcal{S}$-matrix elements, the carrollian amplitudes necessarily transform as correlation functions for the corresponding carrollian conformal fields. We refer the reader to \cite{Nguyen:2025sqk} for further details. 
	
	We now turn to description of asymptotic symmetries within this carrollian picture. We start with the gravitational asymptotic symmetries also known as BMS symmetries \cite{Bondi:1962px,Sachs:1962wk,Sachs:1962zza}, since they most naturally appear as the full set of (globally defined) conformal isometries of~$\scri$. See \cite{Nguyen:2022zgs} and references therein. The BMS group is a semidirect product of the form
	\begin{equation}
	\operatorname{BMS}=\operatorname{SO}(1,3) \ltimes \mathcal{E}_{-1}(S^2)\,.
	\end{equation}
	An element of the \textit{supertranslation} subgroup $\mathcal{E}_{-1}(S^2)$ is a smooth function $T(\vec x)$ over the celestial sphere with weight $\Delta_T=-1$. Among these supertranslations, one recovers the usual time translation as its $\ell=0$ spherical harmonic, and the three spatial translations as its $\ell=1$ spherical harmonics. The Lorentz group $\operatorname{SO}(1,3)$ acts as the group of conformal isometries of the sphere $S^2$, and its action on the function $T(\vec x)$ is specified by saying that the latter transforms as a conformal scalar field of dimension $\Delta_T=-1$. Of interest to us is the action of infinitesimal supertranslations on the carrollian conformal fields~\eqref{Mellin}. Denoting by $Q_T$ the corresponding generator, this is simply given by 
	\begin{equation}
	[Q_T\,, O_{\Delta,J}(\x)]=i \, T(\vec x)\, \partial_u O_{\Delta,J}(\x)\,.
	\end{equation}
	This essentially means that supertranslations couple to the energy of the particles.
	
	The structure of the group of electromagnetic asymptotic symmetries, also called `large' $\operatorname{U}(1)$ symmetries or `phase superrotations', is actually very similar. Strictly speaking, this group is simply $\operatorname{U}(1)_{\text{large}}=\mathcal{E}_{0}(S^2)$, whose elements are again smooth functions $\lambda(\vec x)$ over the celestial sphere. However it is important to recognise that these symmetries are not internal since the Lorentz group has an action on the celestial sphere, such that $\lambda(\vec x)$ transforms as a conformal scalar field of dimension $\Delta_\lambda=0$. We can combine both asymptotic symmetries within the group
	\begin{equation}
	\label{ASG}
	\operatorname{ASG}=\operatorname{SO}(1,3) \ltimes \left[ \mathcal{E}_{-1}(S^2) \times \mathcal{E}_{0}(S^2) \right]\,.
	\end{equation}
	Denoting by $Q_\lambda$ the generator of infinitesimal phase superrotations, its action on a carrollian conformal field $O_{\Delta,J}(\x)$ of electric charge $e$ is given by
	\begin{equation}
	[Q_\lambda\,, O_{\Delta,J}(\x)]=-ie \, \lambda(\vec x)\, O_{\Delta,J}(\x)\,.
	\end{equation}
	The uniform phase rotation ($\lambda=const$) corresponds to the usual $\operatorname{U}(1)_{\text{global}}$ symmetry.
	
	As usual, symmetries yield Ward identities satisfied by correlation functions. In particular, computing the expectation value of $[Q_\lambda\,, O_1(\x_1)\,...\,O_n(\x_n)]$ in the vacuum state $|0 \rangle$ yields the (integrated) Ward identity
	\begin{equation}
	\label{photon Ward identity}
	\begin{split}
	&\langle 0|Q_\lambda\,  O_1(\x_1)\,...\,O_n(\x_n) - O_1(\x_1)\,...\,O_n(\x_n)\, Q_\lambda |0\rangle\\
	&= -i\sum_{a=1}^n e_a\, \lambda(\vec x_a) \langle 0| O_1(\x_1)\,...\,O_n(\x_n) |0 \rangle\,.
	\end{split}
	\end{equation}
	Similarly, the Ward identity resulting from BMS symmetries reads
	\begin{equation}
	\label{graviton Ward identity}
	\begin{split}
	&\langle 0|Q_T\,  O_1(\x_1)\,...\,O_n(\x_n) - O_1(\x_1)\,...\,O_n(\x_n)\, Q_T |0\rangle\\
	&= i\sum_{a=1}^n  T(\vec x_a)\partial_{u_a} \langle 0| O_1(\x_1)\,...\,O_n(\x_n) |0\rangle\,.
	\end{split}
	\end{equation}
	These Ward identities can be used to diagnose spontaneous breaking of asymptotic symmetries. Indeed, if they are unbroken then $Q_{\lambda,T}|0\rangle=0$ and the left-hand sides of \eqref{photon Ward identity}-\eqref{graviton Ward identity} vanish. Conversely, evaluating the right-hand sides and finding a nonzero result implies spontaneous symmetry breaking.  
	
	\section{Spontaneous breaking from soft theorems}
	\label{section 3}
	We now turn to the carrollian description of the soft photon and soft graviton theorems originally derived in~\cite{Weinberg:1965nx}. By comparison to the Ward identities \eqref{photon Ward identity}-\eqref{graviton Ward identity}, we will conclude that they necessarily imply spontaneous symmetry breaking (SSB).    
	
	Let us start by recalling the soft photon theorem, 
	\begin{equation}
	\label{soft photon theorem}
	\begin{split}
	&\lim_{\omega \to 0} \omega\, S_{n+1}\left(p_1\,,...\,, p_n\,; \omega q(\vec y),\varepsilon_i\right)\\
	&=\sum_{a=1}^n e_a\, \frac{p_a \cdot \varepsilon_i(\vec y)}{p_a \cdot q(\vec y)}\, S_n\left(p_1\,,...\,, p_n\right)\,,
	\end{split}
	\end{equation}
	where $\varepsilon_i^\mu(\vec y)$ with $i=1,2$ is one of the two independent polarisation vectors of the soft photon with momentum $\omega q^\mu(\vec y)$. The $p_a$'s are the momenta of the other massless particles involved in the scattering process, and $e_a$'s their electric charges. The polarisation vectors can be explicitly constructed through $\varepsilon_i^\mu(\vec y)=\partial q^\mu(\vec y)/\partial y^i$,
	such that the soft photon theorem \eqref{soft photon theorem} also reads
	\begin{equation}
	\begin{split}
	&\lim_{\omega \to 0} \omega\, S_{n+1}\left(p_1\,,...\,, p_n\,; \omega q(\vec y),\varepsilon_i \right)\\
	&=\sum_{a=1}^n e_a\, \frac{2(y^i-x^i_a)}{|\vec y-\vec x_a|^2}\, S_n\left(p_1\,,...\,, p_n\right)\,.
	\end{split}
	\end{equation}
	Going to the carrollian basis by applying the transform \eqref{Mellin} to the $n$ external legs, we obtain
	\begin{equation}
	\begin{split}
	&\lim_{\omega \to 0} \omega\, \langle O_1(\x_1)\,...\,O_n(\x_n)\, a^\dagger_i(\omega q(\vec y)) \rangle\\
	&=\sum_{a=1}^n e_a\, \frac{2(y^i-x^i_a)}{|\vec y-\vec x_a|^2}\, \langle O_1(\x_1)\,...\,O_n(\x_n) \rangle\,.
	\end{split}
	\end{equation}
	As a last step, we use the $\mathcal{S}$-matrix identity \cite{He:2014cra}
	\begin{equation}
	\lim_{\omega \to 0}\! \omega\, \langle \text{out}| \mathcal{S} a^\dagger_i(\omega q)|\text{in} \rangle\!=-\!\lim_{\omega \to 0}\! \omega\, \langle \text{out}|a_i(\omega q) \mathcal{S} |\text{in} \rangle,
	\end{equation}
	to write it in the form
	\begin{widetext}
		\begin{equation}
		\label{carrollian soft theorem}
		\lim_{\omega \to 0} \omega \left[\langle O_1(\x_1)\,...\,O_n(\x_n)\, a^\dagger_i(\omega q(\vec y)) \rangle- \langle\, a_i(\omega q(\vec y))\, O_1(\x_1)\,...\,O_n(\x_n) \rangle \right] =\sum_{a=1}^n e_a\, \frac{(y^i-x^i_a)}{|\vec y-\vec x_a|^2}\, \langle O_1(\x_1)\,...\,O_n(\x_n) \rangle\,.
		\end{equation}
	\end{widetext}
	At this point we are able to make contact with the Ward identity \eqref{photon Ward identity}. Indeed we see that the right-hand side of \eqref{carrollian soft theorem} reproduces the right-hand of \eqref{photon Ward identity} if we choose the particular symmetry parameter
	\begin{equation}
	\lambda(\vec x;\vec y,\varepsilon_i)=\frac{(y^i-x^i)}{|\vec y-\vec x|^2}\,,
	\end{equation}
	which is determined in terms of the momentum coordinate $\vec y$ and the polarization label $i$ of the soft photon. Since the right-hand is non-zero, \textit{it necessarily implies spontaneous breaking} of the $\operatorname{U}(1)_{\text{large}}$ symmetry within the carrollian field theory framework. We further identify a soft photon state with the shifted vacuum,
	\begin{equation}
	Q_\lambda |0\rangle \equiv i \lim_{\omega \to 0} \omega a^\dagger_i(\omega q(\vec y))|0 \rangle\,,
	\end{equation}
	such that the correspondence between \eqref{photon Ward identity} and \eqref{carrollian soft theorem} is complete. This identification should be viewed as heuristic. Making it mathematically precise will be the subject of the next section. 
	More generally, one can define a family of degenerate vacua 
	\begin{equation}
	\label{degenerate vacua}
	|\lambda\rangle \equiv e^{iQ_\lambda} |0\rangle\,,
	\end{equation}
	where the vacuum $|0\rangle$ serves as an arbitrary reference state. 
	
	Spontaneous breaking of BMS symmetries can be diagnosed in the same way. Starting from the soft graviton theorem
	\begin{equation}
	\label{soft graviton theorem}
	\begin{aligned}
	&\lim_{\omega \to 0} \omega\, S_{n+1}\left(p_1\,,...\,, p_n\,; \omega q(\vec y),\varepsilon_i\right)\\
	&=\kappa \sum_{a=1}^n \frac{(p_a \cdot \varepsilon_i(\vec y))^2}{p_a \cdot q(\vec y)}\, S_n\left(p_1\,,...\,, p_n\right)\\
	&=\kappa \sum_{a=1}^n \eta_a\omega_a\,  \frac{4 (y^i-x^i_a)^2}{|\vec y-\vec x_a|^2}\, S_n\left(p_1\,,...\,, p_n\right)\,,
	\end{aligned}
	\end{equation}
	where $\varepsilon_i^{\mu\nu}(\vec y)=\varepsilon_i^\mu(\vec y)\varepsilon_i^\nu(\vec y)$ is one of two polarisation tensors of the soft graviton with momentum $\omega q^\mu(\vec y)$, and $\kappa=\sqrt{8\pi G}$ the gravitational coupling. Going to the carrollian basis using \eqref{Mellin}, we obtain
	\begin{widetext}
		\begin{equation}
		\label{carrollian soft graviton theorem}
		\lim_{\omega \to 0} \omega\! \left[\langle O_1(\x_1)...O_n(\x_n) a^\dagger_i(\omega q(\vec y)) \rangle\!-\! \langle a_i(\omega q(\vec y)) O_1(\x_1)...O_n(\x_n) \rangle \right]\! 
		=-i\kappa\! \sum_{a=1}^n \frac{2(y^i-x^i_a)^2}{|\vec y-\vec x_a|^2} \partial_{u_a} \langle O_1(\x_1)...O_n(\x_n) \rangle\,.
		\end{equation}
	\end{widetext}
	The right-hand side of \eqref{carrollian soft graviton theorem} reproduces the right-hand of \eqref{graviton Ward identity} if we choose the particular symmetry parameter
	\begin{equation}
	T(\vec x;\vec y,\varepsilon_i)=\frac{2(y^i-x^i)^2}{|\vec y-\vec x|^2}\,,
	\end{equation}
	which is determined in terms of the momentum coordinate $\vec y$ and the polarization label $i$ of the soft graviton. Since the right-hand is non-zero, it  implies spontaneous breaking of BMS supertranslations. To complete the correspondence between \eqref{graviton Ward identity} and \eqref{carrollian soft graviton theorem}, we further identify a soft graviton state with the shifted vacuum,
	\begin{equation}
	\label{shifted vacuum}
	Q_T |0\rangle \equiv \lim_{\omega \to 0} \omega a^\dagger_i(\omega q(\vec y))|0 \rangle\,.
	\end{equation}
	
	We have thus established, using the carrollian field theory description of amplitudes, that Weinberg's soft theorems imply the spontaneous breaking of the asymptotic symmetry group \eqref{ASG}, down to
	\begin{equation}
	\operatorname{ASG} \quad  \stackrel{\text{SSB}}{\longrightarrow} \qquad  \operatorname{ISO}(1,3) \times \operatorname{U}(1)_{\text{global}}\,.
	\end{equation}
	Note that translations and global phase rotations are unbroken. Indeed in that case the right-hand side of \eqref{photon Ward identity}-\eqref{graviton Ward identity} do evaluate to zero on account of charge and momentum conservation.  
	
	\section{Celestial Goldstone particles}
	Let us study the implications of spontaneous symmetry breaking for the spectrum of the theory. Within Minkowskian quantum field theory, spontaneous symmetry breaking typically implies the existence of massless particles in the spectrum. In a carrollian conformal field theory, we show that spontaneous symmetry breaking implies the existence of \textit{zero-momentum} representations in the spectrum. To show this we adapt the argument presented in \cite{Burgess:2020tbq}.
	
	First we write the broken symmetry generators $Q_{\lambda}$ and $Q_T$ as the spatial integrals of some current densities,
	\begin{equation}
	\label{integral charges}
	\begin{split}
	Q_\lambda(u)&=\int_{\Sigma_u} d^2\vec x\, \lambda(\vec x)\, \rho(\x) \,,\\
	Q_T(u)&=\frac{2}{\kappa^2} \int_{\Sigma_u} d^2\vec x\, T(\vec x)\, \mathcal{M}(\x) \,,
	\end{split}
	\end{equation}
	where $\Sigma_u$ is a cut of $\scri$, i.e., a surface of fixed $u$. Here we pause to make contact with the literature. Recalling that $\scri$ happens to be the future null conformal boundary of Minkowski spacetime,  the current density $\rho(\x)$ is identified with the pullback to $\scri$ of the radial component of the electric field \cite{He:2014cra}, while $\mathcal{M}(\x)$ is identified with the BMS supermomentum as given in \cite{Donnay:2022hkf,Agrawal:2023zea} which can further be viewed as a component of the carrollian stress tensor \cite{Donnay:2022aba,Nguyen:2025sqk}. The charges \eqref{integral charges} are the true canonical generators of asymptotic symmetries only when $\Sigma_u$ lies close to spatial infinity, i.e., when $u \to -\infty$. This is the region of spacetime where radiation is absent, and where the current densities are actually conserved. Away from spatial infinity, the current densities rather satisfy the classical evolution equations \cite{Strominger:2017zoo}
	\begin{equation}
	\label{charge loss}
	\partial_u \rho =  -\partial_u \partial^i A_i - e^2 j_u\,,
	\end{equation}
	and \cite{Donnay:2022hkf}
	\begin{equation}
	\label{mass loss}
	\partial_u \mathcal{M} = \frac{1}{4}\left(\partial^i\partial^j N_{ij}+C^{ij}\partial_u N_{ij}\right) -  \kappa^2\, T_{uu}\,.
	\end{equation}
	On the right-hand sides of these equations $A_i(\x)$ and $C_{ij}(\x)$ (with $N_{ij}=\partial_u C_{ij}$) are the carrollian conformal fields carrying photons and gravitons, while $j_u$ and $T_{uu}$ are respectively the flux of electric charge and flux of energy carried away by matter fields.  Equation \eqref{mass loss} is called `Bondi mass loss formula', as it expresses the fact that the total mass of the system contained within spacetime decreases as energy is carried away in the form of radiation. As emphasized previously, the charges $Q_{\lambda,T}(u)$ are canonical generators only in the limit $u\to -\infty$, precisely where the non-conservation terms on the right-hand side of \eqref{charge loss}-\eqref{mass loss} vanish. We will make use of this fact in what follows.
	
	The existence of a spontaneously broken symmetry can be characterised by the existence of a field $O$ that is charged under this symmetry, with symmetry transformation given by
	\begin{equation}
	\label{delta psi}
	[ Q_{\lambda,T}\,, O]=\phi_{\lambda,T}\,,
	\end{equation}
	such that $\phi_{\lambda,T}$ has nonzero expectation value in the vacuum state of reference, 
	\begin{equation}
	\label{order parameter}
	\langle 0|\phi_{\lambda,T}|0\rangle \neq 0\,.
	\end{equation}
	This quantity, usually called order parameter, detects the spontaneous symmetry breaking. Indeed, this is only possible if the vacuum breaks the symmetry,
	\begin{equation}
	Q_{\lambda,T}|0\rangle \neq 0\,.
	\end{equation}
	In the previous section we had considered the operator $O=O_1(\x_1)\,...\,O_n(\x_n)$ to detect this spontaneous symmetry breaking, see \eqref{photon Ward identity}-\eqref{graviton Ward identity}.
	We can then write, 
	\begin{widetext}
		\begin{equation}
		\langle 0|\phi_\lambda |0\rangle=\langle 0|[Q_\lambda\,, O]|0\rangle=\int d^2\vec x\, \lambda(\vec x) \langle 0| [\rho(\x)\,,O]|0\rangle=\sum_n \int d^2\vec x\, \lambda(\vec x) \left[ \langle 0| \rho(\x)|n\rangle \langle n|O|0\rangle -\langle 0| O|n\rangle \langle n|\rho(\x)|0\rangle\right]\,,
		\end{equation}
	\end{widetext}
	where in the second line we have inserted a resolution of the identity on the Hilbert space of the theory, schematically denoted $\mathds{1}=\Sigma_n |n\rangle \langle n|$. Since this quantity is nonzero by \eqref{order parameter}, there must exist at least one particle species $|G\rangle$ such that 
	\begin{equation}
	\label{rho G}
	\langle 0|\rho(\x)|G\rangle \neq 0\,.
	\end{equation}
	We wish to determine the properties of this type of Goldstone particle.
	
	The states $|n\rangle$ are organized according to unitary irreducible representations of the Poincar\'e group. Since we are considering a field theory at $\scri$ the underlying one-particle states are either massless particles, or zero-momentum representations. For massless states one can adopt the momentum parametrisation \eqref{momentum parametrisation}, and denoting $|p\rangle$ the corresponding states, the action of all Poincar\'e generators is explicitly given by
	\cite{Iacobacci:2024laa}
	\begin{equation}
	\tilde P_\mu |p\rangle =p_\mu |p\rangle\,,    
	\end{equation}
	and
	\begin{align}
	P_i\, |p\rangle &=-i \partial_i\, |p \rangle\,,\nonumber\\
	\label{massless rep}
	J_{ij}\, |p\rangle&=-i\left(x_i \partial_j-x_j\partial_i+i \Sigma_{ij} \right) |p\rangle\,,\\
	D\, |p\rangle&=i\left(-\omega \partial_\omega+x^i\partial_i \right)|p\rangle\,,\nonumber\\
	K_i\, |p\rangle &=i\left(-2x_i \omega \partial_\omega+2x_i x^j \partial_j-x^2 \partial_i+2i x^j \Sigma_{ij} \right)|p\rangle\,,\nonumber
	\end{align}
	where $\Sigma_{ij}$ are spin matrices.
	This is the action of momentum and Lorentz generators, respectively, written in a basis most appropriate to $\scri$.
	A zero-momentum representation is one where the momentum generators $\tilde P_\mu$ act trivially, i.e., it is simply a unitary representation of the Lorentz subgroup. The latter are given by the elementary representations $|\Delta,\vec x\rangle$, transforming according to 
	\begin{align}
	P_i\, |\Delta, \vec x \rangle &=-i \partial_i\, |\Delta, \vec x\rangle\,,\nonumber\\
	\label{induced rep}
	J_{ij}\, |\Delta, \vec x\rangle&=-i\left(x_i \partial_j-x_j\partial_i+i \Sigma_{ij} \right) |\Delta, \vec x\rangle\,,\\
	D\, |\Delta, \vec x\rangle&=i\left(\Delta+x^i\partial_i \right)|\Delta, \vec x\rangle\,,\nonumber\\
	K_i\, |\Delta, \vec x\rangle &=i\left(2x_i \Delta+2x_i x^j \partial_j-x^2 \partial_i+2i x^j \Sigma_{ij} \right)|\Delta, \vec x\rangle\,.\nonumber
	\end{align}
	We invite the reader to consult \cite{Sun:2021thf} for a friendly review of elementary representations. Armed with these representations, we can return to \eqref{rho G}. First let us use the carrollian translation operators $H\equiv (\tilde P_0-\tilde P_3)/\sqrt{2}$ and $P_i$ to write \cite{Nguyen:2023vfz,Nguyen:2025sqk}
	\begin{equation}
	\rho(\x)=e^{-iuH} e^{-ix^i P_i} \rho(0) e^{ix^i P_i}e^{iuH}\,,
	\end{equation}
	such that 
	\begin{equation}
	\langle 0|\rho(\x)|G\rangle =\langle 0|e^{-ix^iP_i} \rho(0) e^{ix^i P_i}e^{iuH}|G\rangle\,,
	\end{equation}
	where we have assumed the vacuum to be time-translation invariant, i.e., $H|0\rangle=0$. 
	Then carrollian current conservation near $u \to -\infty$ implies
	\begin{equation}
	\label{current conservation matrix element}
	0=\langle 0| \partial_u \rho(\x)|G\rangle=\partial_u \langle 0|e^{-ix^iP_i}\! \rho(0) e^{ix^i P_i}e^{iuH}|G\rangle,
	\end{equation}
	which can only be satisfied if the Goldstone particle is also time-translation invariant,
	\begin{equation}
	H |G\rangle=0\,.
	\end{equation}
	This only happens if the Goldstone particle belongs to a zero-momentum representation, i.e., $|G\rangle=|\Delta,\vec x\,\rangle_G$. Such a zero-momentum state corresponds to a time-independent carrollian conformal scalar field, or more simply to a $\operatorname{SO}(1,3)$ conformal scalar field on the celestial sphere,
	\begin{equation}
	\label{Goldstone state}
	G_\Delta(\vec x)|0\rangle=|\Delta,\vec x\, \rangle_G\,.
	\end{equation}
	Of course this is in perfect agreement with existing literature which have introduced Goldstone modes as conformal fields on the celestial sphere  \cite{Nande:2017dba,Himwich:2020rro,Nguyen:2021ydb,Donnay:2022hkf,Agrawal:2023zea,Kapec:2021eug,He:2024skc}. Following the discussion around \eqref{ASG}, the scaling dimensions of the electromagnetic and gravitational Goldstone fields are respectively identified to be $\Delta_\lambda=0$ and $\Delta_T=-1$. We note that these values correspond to the discrete unitary series (the shadow dimension $\Delta_s=2-\Delta$ is a positive integer), see \cite{Sun:2021thf} for further details. 
	
	Importantly, the inner product on the Hilbert space corresponding to the unitary discrete series with $\Delta_s=2-\Delta \in \mathbb{N}$ is determined, up to some normalization constant $\mathcal{N}$, by the kernel \cite{Dobrev:1977qv,Sun:2021thf,Schaub:2024rnl}
	\begin{equation}
	\label{kernel}
	{}_G\langle \Delta, \vec x_1|\Delta, \vec x_2 \rangle_G=\mathcal{N}\, |\vec x_{12}|^{-2\Delta} \ln (\mu |\vec x_{12}|)\,,
	\end{equation}
	which is also the two-point function of the Goldstone field according to \eqref{Goldstone state}.
	The inner product for states of the form
	\begin{equation}
	|\psi \rangle_G=\int d^2\vec x\, \psi(\vec x) |\Delta,\vec x\rangle_G\,,
	\end{equation}
	does not actually depend on the arbitray scale $\mu$ as explained in \cite{Dobrev:1977qv,Sun:2021thf,Schaub:2024rnl}. It is very interesting to note that the kernel \eqref{kernel} precisely takes the form of the Goldstone two-point functions identified in \cite{Nande:2017dba,Himwich:2020rro} that control infrared divergences of scattering amplitudes. Here we see that it arises naturally and unavoidably from the nature of the discrete series representations of the Lorentz group, by contrast to other approaches invoking a Jordan block structure characteristic of logarithmic CFTs \cite{Fiorucci:2023lpb,Bissi:2024brf}. 
	
	\section{Discussion}
	While the results presented in this work may seem familiar to some readers, we insist that the derivations given here are entirely field theoretic. Without ever invoking spacetime fields, but rather relying on carrollian fields and the underlying $\operatorname{ISO}(1,3)$ quantum states, we have shown that soft theorems \textit{imply} spontaneous breaking of asymptotic symmetries as defined in \eqref{ASG}, and that the corresponding Goldstone particles belong to zero-momentum representations, i.e., they are quantum conformal fields on the celestial sphere. Moreover, the logarithmic two-point functions introduced such as to account for infrared divergences of scattering amplitudes \cite{Nande:2017dba,Himwich:2020rro} naturally follow from the inner product that the discrete series representations of the Lorentz group come equipped with.
	
	This new perspective on soft theorems potentially has deep implications for scattering theory with photons and gravitons. Indeed let us recall that the conventional treatment relying on Feynman diagrams suffers from infrared divergences that are directly related to the soft behavior seen in \eqref{soft photon theorem} and \eqref{soft graviton theorem}  \cite{Weinberg:1965nx}. In fact, this $\omega^{-1}$ behavior is simply \textit{inconsistent} with the Hilbert space of photons and gravitons as discussed for example in \cite{Ashtekar:2018lor,Satishchandran:2019pyc}. The norm of a photon/graviton state $|\psi \rangle$ of the form
	\begin{equation}
	|\psi \rangle =\sum_{i=+,-} \int d^2\vec x\int_0^\infty d\omega\, \psi_i(\omega,\vec x) |p(\omega,\vec x)\rangle_i\,,
	\end{equation}
	is given by
	\begin{equation}
	|| \psi ||^2=\sum_{i=+,-}\int d^2\vec x\, \int_0^\infty d\omega\, \omega\, |\psi_i(\omega,\vec x)|^2\,,
	\end{equation}
	and a behavior $\psi_i(\omega) \sim \omega^{-1}$ yields a logarithmic divergence from the lower bound of the integral. What the picture given in this paper suggests is to associate Weinberg's soft pole not with conventional photon or graviton states of vanishing energy, but rather with Goldstone states that come with their own Hilbert space. Indeed, equation \eqref{rho G} essentially implies that the broken currents are built from the Goldstone fields, such that the shifted vacuum \eqref{shifted vacuum}, formally identified with the ill-defined soft photon or soft graviton, is actually a Goldstone state. To some extent what we suggest here is already implicit in the treatment of infrared divergences within celestial holography \cite{Nande:2017dba,Himwich:2020rro,Nguyen:2021ydb,Donnay:2022hkf,Agrawal:2023zea,Kapec:2021eug,He:2024skc}, although the proposed description may vary. An exciting prospect would be the formulation of scattering amplitudes that are free of infrared divergences using the framework described here, with possible connections to other approaches \cite{Kulish:1970ut,Zwanziger:1973if,Zwanziger:1973mp,Zwanziger:1974jz,Zwanziger:1974ka,Zwanziger:1976vv,Jakob:1990zi,Forde:2003jt,Ware:2013zja,Kapec:2017tkm,Hannesdottir:2019umk,Hannesdottir:2019opa,He:2020ifr,Kapec:2022hih,Prabhu:2022zcr,Prabhu:2024lmg,Prabhu:2024zwl,Bekaert:2024jxs}.
	
	Finally, let us note that spontaneous symmetry breaking usually comes with interesting constraints on low-energy dynamics and physical observables \cite{Burgess:2020tbq}, of which Adler's zero is a perfect example \cite{Adler:1964um}. Interestingly, the carrollian counterpart of Adler's zero should result in a constraint on the carrollian operator product expansion that has been the subject of recent investigation~\cite{Nguyen:2025sqk}. To see this, let us consider a correlation function containing one insertion of the broken current $\langle  \rho(\x_0) O_1(\x_1) O_2(\x_2)\,...\,\rangle$.
	We can use the operator product expansion 
	\begin{equation}
	\label{OPE}
	O_1(\x_1)O_2(\x_2) \sim \sum_{O_3} \left[ f_{123}(\x_{12})\, O_3(\x_2)+subl.\,\right]\,,
	\end{equation}
	where the sum runs over primary operators $O_3$, and where subleading terms feature their derivatives. See \cite{Nguyen:2025sqk} for further details. Among these primary operators is the Goldstone field $G(\vec x)$, such that
	\begin{widetext}
		\begin{equation}
		\begin{split}
		\langle  \rho(\x_0) O_1(\x_1) O_2(\x_2)\,...\, \rangle & \sim \left[ f_{12G}(\x_{12})\langle  \rho(\x_0) G(\vec x_2)\,...\, \rangle+subl.\,\right]+R(\x_0,\x_1,\x_2,...)\,,
		\end{split}
		\end{equation}
	\end{widetext}
	where the remainder $R(\x_0,\x_1,\x_2,...)$ accounts for the contributions of all other primaries and descendants. Provided we are able to determine the correlator $\langle  \rho(\x_0) G(\vec x_2)...\, \rangle$, using \eqref{charge loss} we can potentially infer the time-derivative $\partial_{u_0} R(\x_0,\x_1,\x_2,...)$ of the remainder. This potentially places interesting constraints on the carrollian operator product expansion, and further motivates the development of a carrollian conformal bootstrap as envisioned in \cite{Nguyen:2025sqk}. 
	
	\section*{Acknowledgments}
	We thank Jakob Salzer for useful discussions. The work of KN is supported by a Postdoctoral
	Research Fellowship granted by the FNRS (Belgium). SA is partially supported by INFN Iniziativa Specifica ST\&FI and by the European Research Council (ERC) Project 101076737 -- CeleBH. Views and opinions expressed are however those of the author only and do not necessarily reflect those of the European Union or the European Research Council. Neither the European Union nor the granting authority can be held responsible for them.
	
	\bibliography{bibl}
	\bibliographystyle{JHEP}  
	
\end{document}